\documentclass[11pt]{article}
\usepackage[OT2, T1]{fontenc}
\usepackage[english]{babel}
\usepackage{authblk}
\usepackage{epsf}
\usepackage[dvips]{graphics,graphicx}
\usepackage{cite}
\usepackage{enumerate}
\usepackage[normalem]{ulem}
\usepackage{amsmath,amssymb,bm,color,mathrsfs,wasysym}
\usepackage{epsfig}
\usepackage{amsfonts}
\usepackage{cancel}
\usepackage{float}
\usepackage{perpage}
\usepackage{setspace}
\usepackage{nameref}
\usepackage{hyperref}
\hypersetup{colorlinks}
\usepackage{xcolor}
\usepackage[lmargin=3cm,rmargin=3cm,tmargin=2cm,bmargin=2cm,marginpar=2cm ,reversemp]{geometry}

\title{The Effect of Defects on Magnetic Droplet Nucleation}

\begin{document}

    \author[1]{Federico Ettori
    \thanks{federico.ettori@polimi.it}}
    \author[1,2]{Timothy J. Sluckin
    \thanks{t.j.sluckin@soton.ac.uk}}
    \author[1]{Paolo Biscari
    \thanks{paolo.biscari@polimi.it}}
	
    \affil[1]{\small Department of Physics, Politecnico di Milano, Piazza Leonardo da Vinci 32, 20133 Milan, Italy}
    \affil[2]{\small School of Mathematical Sciences, University of Southampton, University Road, Highfield, Southampton, SO17 1BJ, UK}
    \date{}
    
\maketitle

\begin{abstract}
    \noindent Defects and impurities strongly affect the timing and the character of the (re)ordering or disordering transitions of thermodynamic systems captured in metastable states. In this paper we analyze the case of two-dimensional magnetic systems. We adapt the classical JMAK theory to account for the effects of defects on the free energy barriers, the critical droplet area and the associated metastable time. The resulting predictions are successfully tested against the Monte-Carlo simulations performed by adopting Glauber dynamics, to obtain reliable time-dependent results during the out-of-equilibrium transformations. We also focus on finite-size effects, and study how the spinodal line (separating the single-droplet from the multi-droplet regime) depends on the system size, the defect fraction, and the external field.
\end{abstract}
  
\section{Introduction}\label{sec:I}
	
	Macroscopic systems exhibit a large variety of phase transitions when subject to continuous variations of external control parameters, such as temperature, magnetic field, pressure, or chemical potential. The nature and timing of the nucleation and growth processes associated with the transition strongly depend on a number of factors, including the nature of the transition and the presence of defects and/or impurities in the system. Ehrenfest \cite{ehren,jaeger} first classified phase transitions. In first-order transitions the system jumps discontinuously to a different free energy branch, with a consequent discontinuous jump in the free energy gradient. Higher order transitions were then identified through the order of the discontinuous derivatives of the free energy, though now we know that this is an over-simplified cartoon, and we talk of first-order and continuous transitions.
	
	In the original work it was thought that there was no singularity in the free energy at a first-order transition and that the free energy curve could be analytically continued into a \emph{metastable} region until well beyond the point at which the transition should have taken place. Eventually the system reaches a point at which a susceptibility diverges, and the metastable state become unstable with respect to fluctuations of any sort. This is known as the spinodal line (\emph{line}, because one can draw a set of such points in the full phase space). The fluctuations then grow, and the process is known as spinodal decomposition. In a liquid-gas system, the signature of the spinodal line is the instability with respect to density fluctuations. The system then develops spontaneous density fluctuations and decays into regions of higher and lower density, a process which stops when the densities become equal to the relevant equilibrium liquid and gas densities. Droplets and/or bubbles form; at later times the droplets coagulate, eventually triggering a full phase separation in which the liquid sits at the bottom of the sample and the vapour at the top.
	
	This theoretical picture of metastability is matched experimentally, even though we now know that actually there is an essential singularity in the gradient at the phase transition. All mean field theories, computer simulations and real systems agree in that this metastable phase has a real existence, incidentally in the process demonstrating that too rigid an insistence on a knowledge of equilibrium properties can sometimes be self-defeating in statistical mechanical studies. However, usually thermodynamic systems do not reach the spinodal line and some other process intervenes beforehand causing phase separation and sending the system toward a phase coexistence of true equilibrium phases. What is required in such cases is so-called \emph{nucleation} of the new phase.
	
	The process of nucleation and subsequent growth has thus been the focus of much study over the years. It is also the main focus of the present study,  in which we are particularly concerned with the effects of defects and impurities on nucleation and droplet growth. The nucleation involves the formation of a \emph{nucleus} of the new phase, which then grows (depending on various conservation laws) to invade the whole of the rest of the system, or alternatively enforces phase separation into two coexisting equilibria. Examples might be a new solid phase (say, martensite) invading another which had previously been stable (e.g.\ austenite), a new magnetic phase (say, spin up) invading a formerly stable spin-down phase, or a homogeneous metallic A-B alloy separating into two coexisting  alloys, one A-rich and the other B-rich.
	How this nucleation occurs is itself a subject of considerable study. Textbooks usually draw a distinction between \emph{homogeneous} and \emph{heterogeneous} nucleation.
	
	In homogeneous nucleation droplets of the new phase form (and then usually decay) as fluctuations around the original phase. Droplets in this condition are called subcritical. To reach the size at which they would grow spontaneously (supercritical droplets) involves a spontaneous fluctuation with a free energy large enough to overcome the energy barrier $\Delta F_\text{barr}$. When this is the case the classical Johnson-Mehl-Avrami-Kolmogorov (JMAK) theory \cite{fanfoni,kolmo,avra,john} is expected to capture the main features of the transition. If the system is kept at constant temperature $T$, the probability of fluctuations of the proper size is proportional to $\exp(-\Delta F_\text{barr}/(k_B T))$, with $k_\text{B}$ the Boltzmann constant. As a consequence, the process of developing such a fluctuation is Poisson-like, with the expected time thus proportional to $\exp(+\Delta F_\text{barr}/(k_B T))$, possibly with a complex prefactor in front.
	
	By contrast, heterogeneous nucleation involves requires either \emph{large} (i.e., of a dimension much larger than molecular dimensions) or \emph{many} impurities to seed the transition process. The homo/heterogenous character of a phase transition deeply influences its character as well as the \emph{metastable lifetime}, that is, the mean decay time of a metastable state. When nucleation is originated by sufficiently many seeds, the phase transition process is much more rapid and predictable, in the sense that the standard deviation of the metastable lifetime is reduced. Clearly, the number of nucleation seeds depends on the size of the system itself: the larger the system the easier to nucleate transition droplets. For this reason, the term spinodal line has also been used (see e.g.~\cite{Rikvold1994}), and will here be used, to identify the minimum (finite) size of a system which exhibits heterogeneous nucleation.
	
	The full phase transformation thus involves several distinct stages. Firstly there is an early-stage stochastic nucleation process. This is followed by a deterministic phase during which the critical droplet is growing. Finally there is a late stage stochastic phase, during which the different droplets amalgamate. The dynamics might be expected to be different depending on conservation laws which dictate the final state equilibrium (is it phase transformation, or merely phase separation). Until the advent of computer simulation, these processes were difficult to examine in detail.  Even with computer simulations, the sizes involved in the case of heterogeneous nucleation, not to mention the influence of the finite size of the simulation box, present significant difficulties. The present study seeks to alleviate some of the simulation difficulties by studying a very simple model. We seek to extend understanding of nucleation processes by considering a case which encompasses both homogeneous and heterogeneous nucleation. In our study, the impurities are molecule-sized rather than, as in the case of classical heterogeneous nucleation, colloid-particle-sized. Our simulation uses a two-dimensional Ising model, which, of course, possesses a distinguished lineage in the history of Statistical Mechanics \cite{ons44,kada66}. Spins may be either up or down, are coupled to their neighbours on a lattice.
	
	The precise aim of the present study is to analyze how the presence of impurities influences the relevant phase transitions, with particular focus on nucleation and growth of critical droplets, and the homo/heterogenous character of the transition itself. Common approaches for the modelling of imperfections include the classical Random-Bond \cite{edwand,Fytas2010} and Random-Field Ising Models, where typically Gaussian-distributed bonds and/or quenched local fields \cite{75imry,naskar, Fytas2013} embed the randomness, and diluted Ising lattices \cite{sear} where non magnetic impurities are mixed with magnetic structure. The presence of quenched disorder requires a proper treatment of finite-size effects, including sample-to-sample fluctuations and the possible lack of self-averaging \cite{Fytas2006}. In our study we do this in two ways. First, we choose carefully the number of different samples with the same disorder statistics -- replicas --  over which any physical quantity must be averaged. Second,  we focus on the different regimes that might be present is systems of different sizes (see section \ref{subsec:SL}).
	
	The present paper is organized as follows. In the next section we specify the modelling of defects, and describe the Monte-Carlo algorithm used in the numerical simulations. In Section 3 we focus on the nucleation process. We derive a theoretical prediction of the influence of impurities on the critical droplet size, and the corresponding free energy barrier, and test the prediction against the outcomes of the simulations. In Section \ref{sec:ML} we focus on the homo/heterogenous character of the transition in the presence of defects, and draw conclusions about the spinodal line. A careful analysis of the simulations allows us to analyze also the finite-size effects. A concluding section summarizes and discusses the main outcomes of the present study.
	
	\section{Model and algorithm}
	
	We consider a ferromagnetic Ising Model \cite{Ising1925} occupying a $L\times L$ two-dimensional square lattice, in which periodic boundary conditions are enforced. The system Hamiltonian takes the standard form
	\begin{equation}
		\mathcal{H}[\,\mathrm{s}\,] = -J\sum_{\langle i,j \rangle}s_i s_j - h_\text{ext} \sum_i s_i,
	\end{equation}
	in which $J$ represents the coupling interaction between neighboring spins and $h_\text{ext}$ the magnetic field applied to the system. All the simulations discussed in the present paper adapt the $n$-fold way algorithm first introduced by Bortz et al. \cite{Bortz1975}. In its original 2D version, each of the $N=L^2$ spins is assigned to one among $n=10$ classes, based on their orientation and the number of positively oriented neighbors. This allows us to easily monitor the spins which are most/less likely to modify their state, and therefore to build a rejection-free Monte-Carlo algorithm.
	
	Defects are modelled as fixed spins which are not allowed to modify their orientation during the evolution of the system. In a finite temperature Monte-Carlo simulation the defect-flipping probability cannot be ruled out. Therefore here we are in fact assuming that the characteristic defect-flipping time is larger than the longest simulation time considered (the metastable lifetime analyzed in Sect.~\ref{sec:ML}). The presence of defects does not influence the efficiency of the Monte-Carlo simulation. Ergodicity is ensured provided we restrict the phase space to the free spins. Similarly, the detailed-balance condition is not affected by the inclusion of defects. In the implementation of the algorithm, the defects are all assigned to an additional 11-th class, whose transition probability is held fixed at zero.
	
	In order to understand how the quenched disorder influences the physical properties of the system we study in parallel several different realizations of the system with similar quenched-disorder characteristics. In all realizations of the same system we introduce the same number of defects of positive and negative defects, so to study neutral samples and therefore reduce the sample-to-sample variation \cite{Landau2014}. We parameterize the number of defects through their total fraction $f$ in the system. Therefore, when a fraction $f$ of defects is reported, it means that there are precisely $\lfloor fL^2/2 \rfloor$ quenched defects of each sign distributed at random throughout the system.
	
	\subsection{Defects as random fields}\label{subsec:DRF}
	
	It is known that in 2D the perfect Ising model sustains a low-temperature ferromagnetic phase, characterized by long-range order \cite{ons44,GPU10}. The addition of defect sites introduces quenched randomness and possible frustration. Although randomness and frustration are known to be two key ingredients leading to spin-glass phases \cite{11thomas}, we now show that in the thermodynamic limit no such behavior should be expected.
	
	The presence of defects can be interpreted as the effect of a peculiar type random-field distribution. To be more precise, let $D^{+}$ (resp.\ $D^{-}$) denote the set of defects with fixed positive (negative) orientation, and consider the following random-field distribution on the entire system
	\begin{equation}
		h_{\text{RF},i} = \begin{cases}
			+h_\text{RF} &\text{if $i \in D^{+}$},
			\\
			-h_\text{RF} &\text{if $i \in D^{-}$},
			\\
			\phantom{-}0 &\text{otherwise}.
		\end{cases}
	\end{equation}
	In the $h_\text{RF} \gg J$ regime, any reversal of the selected spins is prevented at any finite (non-zero and non-infinite) temperature, so those spins will effectively behave as \emph{defects}. How large should $h_\text{RF}$ be taken depends on the chosen temperature. It is well know that a random-field Ising model in thermal equilibrium and in the thermodynamical limit, no spin glass phase can be observed \cite{11ker,15chat}. Moreover, no ordered phase can survive in the two-dimensional random-field Ising model \cite{70lark,75imry}. As a result, in the thermodynamic limit, no ferromagnetic nor spin-glass phases are to be expected. What we can and we do observe instead are \emph{pseudo}-phases \cite{11timo} in finite systems, where domain clusterization and finite-size effects generate pseudo-ferro or pseudo-glassy phases, depending on the temperature and defect density. We postpone to a later study the report of the characterization of such pseudo-phases.
	
	\subsection{Dynamic Monte-Carlo algorithm}\label{subsec:DMCA}
	
	The simulations presented in the present work were performed by using the so-called $n$-fold Monte-Carlo algorithm \cite{Bortz1975}, subject to  Glauber dynamics \cite{Glauber1963}. We now discuss how this algorithm is adapted so as to describe the out-of-equilibrium response of magnetic systems in the presence of defects.
	
	The primary quantities of interest in our simulations concern magnitudes and time scales  associated with droplet formation and growth processes.
	More generally, it is thus necessary to characterise the out-of-equilibrium and dynamic response of the system to perturbation. To implement the Glauber dynamics, we associate with each possible single spin flip $s_i\to-s_i$ the transition probability rate
	\begin{equation}
		w_i[\,\mathrm{s}\,] = \frac{1}{2\alpha}\big(1-s_i \tanh\beta\, h_i[\,\mathrm{s}\,]\big),
	\end{equation}
	where $\alpha$ is a microscopic characteristic time, and $h_i = h_\text{ext} + \sum _j J_{ij}s_j$ is the local field acting on the $i$-th spin.
	
	We recall that in our model the transition probabilities associated with the defects are set equal to zero.  In an underlying real physical system,  this will not in general be rigorously true. Such a  system would rather be composed of two species with two very different microscopic characteristic times $\alpha \ll \alpha_\text{def}$. Our key approximation involves taking the limit $\alpha_\text{def}/\alpha \to +\infty$.  This condition is rather strong and may be weakened in future studies. With this choice, we ensure that at any specific time it is much more probable that a normal spin flips rather than a defect. As noted above in our discussion of the basic model, our simulations also rely on yet another asymptotic limit, $\alpha_\text{def} / \tau \to \infty$.  Here $\tau$ represents the longest time in our simulations.  In Section~\ref{sec:ML} this  is labeled  the metastable lifetime. This second limit ensures that no defect can possibly flip during the numerical experiments.
	
	The sum $w_\text{T}=\sum_i w_i$ provides the global transition rate for the entire system.
	The interaction energy $J$ and the characteristic time $\alpha$ are chosen as units for  energy and time, respectively. Then for each Monte-Carlo step,  two operations are required. These are:
	\begin{enumerate}
		\item \label{step1} Identify  the associated time interval $\Delta t$.   This involves extracting a random number from a Poisson distribution with parameter $w_\text{T}^{-1}$.
		\item  Perform  the move which occurs over this time interval $\Delta t$.   A specific spin $i$ to be flipped  is chosen with  probability proportional to $w_i$.   To enable this choice to be made, we use   the Bortz  $n$-fold algorithm  \cite{Bortz1975}.
	\end{enumerate}

	\section{Droplets and defects}
	\label{sec:DD}
	
	We now turn our attention to droplet formation and growth in the presence of defects. We first adapt some theoretical predictions to account for the presence of defects. This part of our analysis relies on the Droplet Theory, adapted to a defect-free magnetic system as in \cite{Rikvold1994} and then report and discuss the results of a number of simulations which help us understanding the proper behaviour of a magnetic system out of equilibrium.
	
	\subsection{Domains and free energy balance}\label{subsec:DFEB}
	
	We consider an initially magnetized system, in which all spins,  with the exception of  the negative defects,  are aligned parallel, and equal to +1. We apply a negative field $h_\text{ext}=-h$, with $h>0$, and study the reversal transition in which droplets of negative spins are expected to form and grow. We identify a droplet as a connected domain, possibly including $n^+$ positive and $n^-$ negative defects, in which all the free spins have already reversed their sign. Positive defecs are counted as part of a negative droplet only if internal to it, that is if it is surrounded by four already reversed spins. We notice that the positive defects are aligned with the initial orientation of all remaining spins, while the negative ones are aligned with the external magnetic field, and act as seeds for droplet formation.
	
	To characterize a droplet we introduce the following parameters:
	
	\begin{itemize}
		\item The area $A$ counts all the spins in the droplet, including the defects.
		\item The perimeter $p$ represents the number of spins defining the droplet boundary. We also introduce a related non-dimensional quantity, which we denote as the \emph{geometric parameter} $\lambda=4\pi A/p^2$. For any simply connected droplet $\lambda$ ranges in the interval $(0,1]$. In a circular domain $\lambda=1$,  in a square droplet it drops to $\pi/4$, whereas for very thin domains, or those with very jagged boundaries, $\lambda \to 0$.
		\item The defect imbalance $\mu=(n^--n^+)/(n^-+n^+)$. By definition, the imbalance $\mu$ is bounded between $-1$ (all enclosed defects are positive) and +1 (all negative defects). It is expected to vanish when the droplet becomes large, as the defects eventually balance. If we focus on the droplet which nucleates the reversal process, we expect $\mu$ to be close to +1, as this domain will most probably be located in a region where predominantly negative defects were located.
		\item The total number of defects included in the domain, $n_\text{T}=n^-+n^+$. Defects are randomly distributed in space.  The Central Limit Theorem  implies that for  sufficiently large  droplets the defects can be regarded as being   homogeneously distributed, with the consequence that
		\begin{subequations}
		\begin{equation}
			n_\text{T}\simeq f A \qquad \text{(homogeneous defect distribution)}.
			\label{Eq:N_t}
		\end{equation}
		However, nucleating droplets  may choose regions where there are more defects, possibly including a gross defect imbalance ($\mu \sim 1$).  For these droplets, eq.~\eqref{Eq:N_t} is not expected to hold any longer.  We introduce the quantity $\nu(f)$  to describe this phenomenon, where
		\begin{equation}
			n_\text{T}=\nu(f) A.
			\label{Eq:nuN_t}
		\end{equation}
		\end{subequations}
		The function $\nu(f)$ denotes the fraction of defects \emph{within the specific droplets under consideration}. Clearly, for sufficiently large droplets, the homogeneous assumption applies and $\nu(f)\approx f$.  On the other hand, in the nucleating droplets, we expect that $\nu(f) \ll f$ for small $f$.
		
	
	\end{itemize}	

\begin{figure}
		\centering
		\includegraphics[width=6cm]{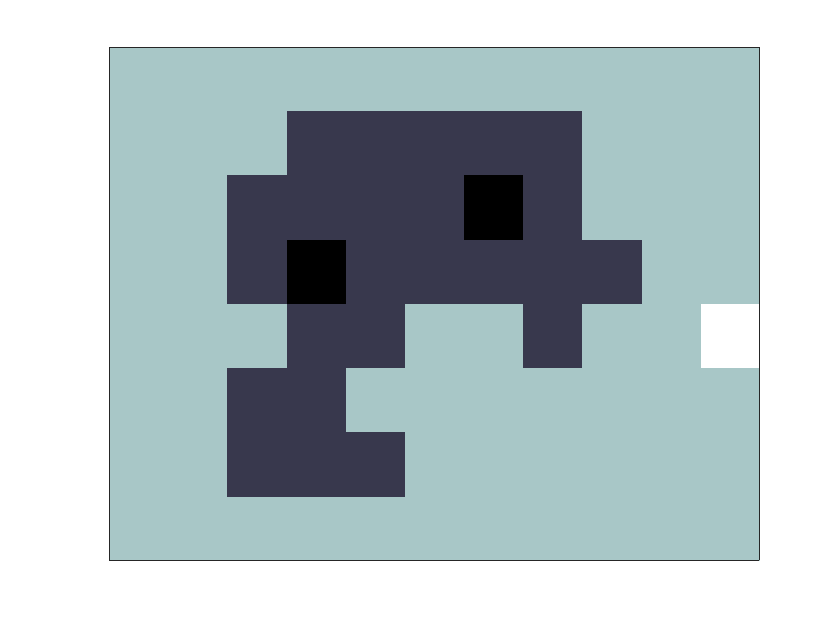}
		\includegraphics[width=6cm]{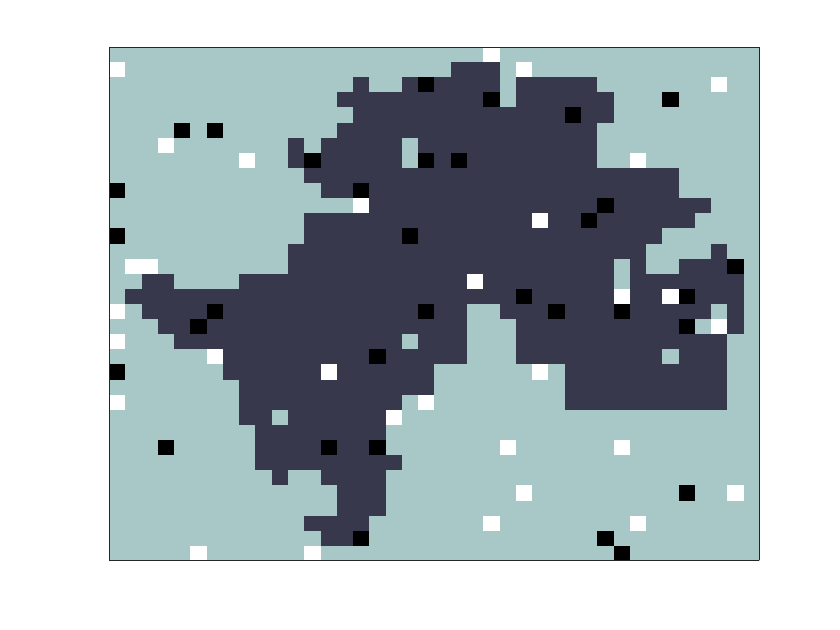}
		\caption{Zoom into a $100 \times 100$ system after the creation of some droplets. The fraction of defects is fixed at $f=4\%$ in both cases. Color code: black and white spins respectively identify negative and positive defects; dark grey and light grey spins are variable spins whose value is resp.\ -1 and +1. The droplet areas are resp.\ 26 and 567 spins. The right droplet can be considered as \emph{large} as it already exceeds the critical nucleation threshold, but it still occupies about 5.7\% of the system. The geometric factor $\lambda$ defined in the text is resp.\ equal to 0.32 for the smaller droplet and 0.13 for the larger one.}
		\label{fig:Droplets}
	\end{figure}

	Figure~\ref{fig:Droplets} shows two droplets with different size. Both are extracted from a simulation of a $100\times 100$ system with a defect density $f=4\%$ (therefore, in average there is a defect every 25 spins). The left panel shows a small ($A=26$), almost critical, droplet which has been originated by two negative defects (already aligned with the external field). Since it encloses $n_\text{T}=2$ defects, it has $\nu=0.08$, larger than the homogeneous value 0.04. By contrast, the larger ($A=567$ spins) droplet in the right plot has $n_\text{T}=29$, so that $\nu=0.05$. Also the parameter $\mu$ behaves as expected, with $\mu=1$ (all negative defects) in the left panel, and $\mu=0.7$ in the right droplet, more balanced with the inclusion also of positive defects.
	
	By adapting \cite{Rikvold1994}, we compare the free energy of the configuration in which the system possesses the spontaneous magnetization $m_\text{s}$, against the generation of a droplet. This involves surface and bulk contributions.
	
	We first consider the surface terms. If we let $\sigma$ be the surface tension, the cost associated with the formation of the droplet boundary is given by
	\begin{equation}\label{eq:fsurf}
		\Delta F_\text{surf} = \sigma \big(p+ 4(n^{+} - n^{-}) \big).
	\end{equation}
	This estimate involves three contributions. The  term  proportional to $p$ evaluates the cost of  creating  the external boundary of the droplet length $p$. In addition, there is an energy cost around each of the $n^+$ positive defects, cancelled by  the benefit associated with the $n^-$ negative defects.
	
	The bulk contribution  is given by
	\begin{equation}
		\Delta F_\text{bulk} = -2m_\text{s}\,h\,\big(A -( n^{+} + n^{-})\big).
	\end{equation}
	This estimate is proportional to the effective area of the droplet $A'=A -n_\text{T}$. The effective area is reduced by the number of defects enclosed in the area. The defect spins are invariant and therefore do not contribute to the energy gain.
	
	The total free energy associated with the creation of a droplet is thus
	\begin{equation}\label{eq:enbar}
		\Delta F = \Delta F_\text{surf}+\Delta F_\text{bulk}=\sigma \left(\sqrt\frac{4\pi A}{\lambda}- 4\mu\,\nu(f)A\right)-2m_\text{s} h A \big(1-\nu(f)\big).
	\end{equation}
	The maximum value of the expression \eqref{eq:enbar} quantifies the height of the free energy barrier that must be overcome to trigger the reversal process. The height of the barrier will certainly decrease with the intensity $h$ of the applied field, but the estimate still depends on a number of other elements, both of geometric origin ($\lambda$) and related to the defect distribution within the droplet ($\mu$, $\nu$). The following probabilistic analysis will help us in identifying the expected behavior for some of these parameters, allowing us to understand how the critical droplet size and the free energy barrier are expected to depend on the external field and the defect density.
	
	\subsection{Critical size and growth velocity}
	\label{subsec:CSGV}
	
	When the magnetic field is reversed, the droplets which are most likely to nucleate are located in those regions with the most negative and the fewest positive defects. For quantitative estimates, it is therefore necessary to study the probability of finding such domains when a defect fraction $f$ is quenched in random positions in the system. In the appendix we also derive the relevant calculations to estimate the probability of finding specific numbers $(n^+,n^-)$ of positive and negative defects in an area $A$, provided that each site has a probability $f/2$ of hosting a positive and a similar probability to host a negative defect.
	
	Let us consider a square system composed of $N=L^2$ spins. If we aim at locating a critical droplet of area $A$ in such a system, there will be $K=N/A$ independent locations where the domain can be placed. Each of these locations will have a different distribution of positive and negative defects, and we are interested in estimating $\Delta n_\text{opt}$, the most probable largest defect imbalance $\Delta n=n^--n^+$ among these $K$ different options. In the appendix (see eq.~\eqref{eq:delnopt}) we derive an expression for such a probability as a function of $K$, $A$, and $f$. By applying Bayesian probability considerations we also derive there the most probable number of defects $n_\text{T,opt}$ hosted in the optimal domain location corresponding to $\Delta n_\text{opt}$.
	
	Figure~\ref{fig:estimate} shows the results obtained by applying the calculations in the appendix to the parameter values appropriate to understand the simulations discussed below.
These calculations consider different defect fractions $f$; the domain area $A$ is set equal to a typical size corresponding to the critical droplets ($A\simeq 45$, see below), and the system size is set equal to $N=10^4$.

With these parameters, the upper panel shows that the optimal imbalance $\Delta n_\text{opt}$ depends algebraically on $f$, as the theoretical prediction fits rather well to a straight line in a log-log plot. The linear fit in the log-log plot is shown in the figure for better comparison. The lower panel compares the same prediction $\Delta n_\text{opt}$ (in blue) with the corresponding expectation for the total number of defects $n_\text{T,opt}$ (orange line, above the blue one). The inset shows that when the defect fraction $f$ is small enough ($f\lesssim 0.1$) the optimal domain contains very few positive defects, so that the total number of defects in that domain is only slightly larger than the imbalance itself. Equivalently, when  defects are sufficiently sparse, it is possible to find a location (among the $K$ available positions) where (almost) all the defects are negative, and therefore $n_\text{T}\simeq\Delta n_\text{opt}$, corresponding to $\mu\simeq 1$.

\begin{figure}
\centering
		\centerline{\includegraphics[width=0.8\linewidth]{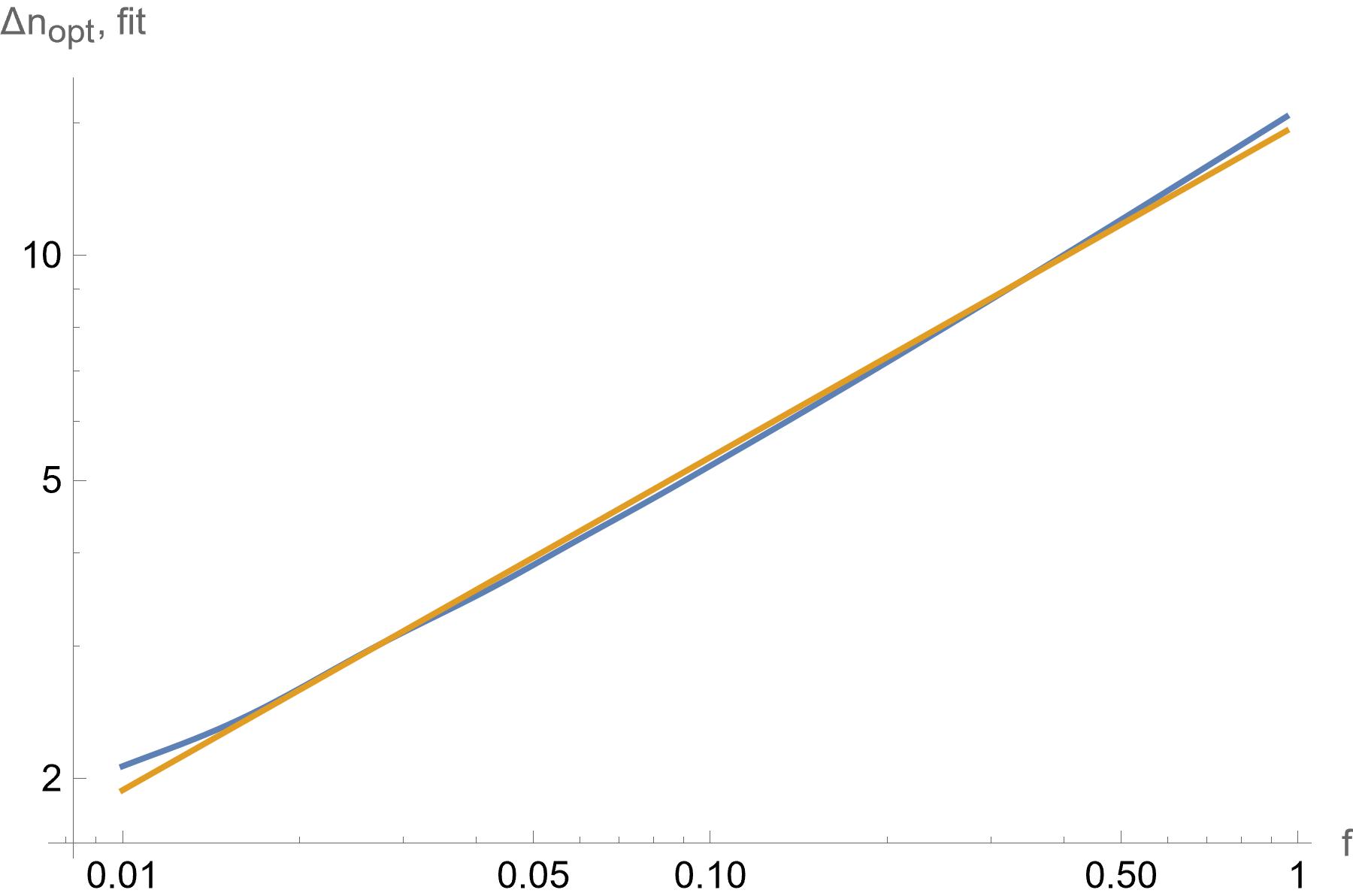}}
		\ \\ \ \\
		\centerline{\includegraphics[width=0.8\linewidth]{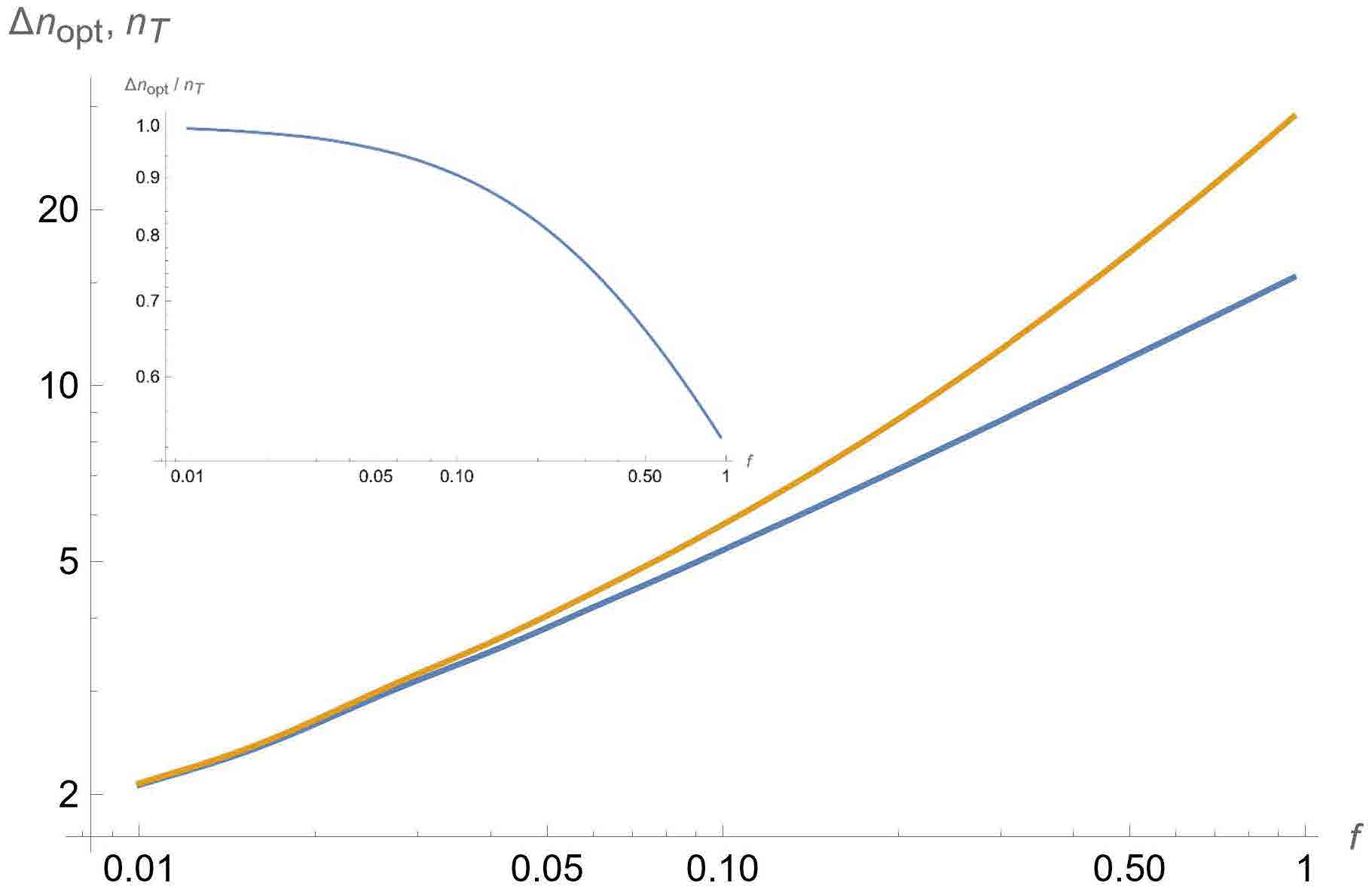}}
		\caption{\textbf{Upper panel}: Probabilistic predictions for a domain of area $A=45$. The optimal value for the defect unbalance $\Delta n$
(blue curve) is calculated by replicating the defect generation $K$ times ($K=N/A$, with $N=10^4$) and keeping the domain with the largest imbalance. The orange line shows that the theoretical prediction fits to the algebraic functional dependence $\Delta n_\text{opt}=\nu_\text{opt}(f) A$, with $\nu_\text{opt}(f)=0.5 f^{0.5}$. \newline
		\textbf{Lower panel}: Total number of defects ($n_\text{T}$, orange, upper curve) corresponding to the optimal imbalance $\Delta n_\text{opt}$ (blue curve, identical as in the upper panel). In the inset one sees clearly that for small values of $f$, the ratio $\mu_\text{opt}=\Delta n_\text{opt}/n_\text{T,opt}$ can be approximated by 1.}
		\label{fig:estimate}
	\end{figure}
			
	By using the above theoretical estimates we may reduce the number of free parameters in the prediction eq.~\eqref{eq:enbar} by setting
	$\mu_\text{opt}\approx 1$ and $\nu_\text{opt}(f)=0.5 f^\alpha$, with $\alpha\approx0.5$. We note the distinction between the result of eq.~\eqref{Eq:nuN_t}, in which $\nu(f) = f$ (i.e. $\alpha=1$)  and this result:
\begin{equation}
		\nu(f) \sim f^{0.5}
		\label{Eq:CriticalNu}
	\end{equation}	
	However, these results apply to two different situations, in the following way.
	
	 The estimates for $(\mu,\alpha)$  considered here are a function of the domain size $A$, and more precisely on the ratio between $A$ and  the global system size $N$.   The regions under consideration correspond to critical droplets.   Positive and negative defects in principle occur with the same  probability.  For small clusters, there will be fluctuations, giving rise to  a higher (or lower)  imbalance between the number of positive and negative defects, and a higher (or lower) local density of defects.  The critical droplets are chosen selectively \emph{exactly because} they correspond locally to regions of higher imbalance and higher local defect density.
	
	  On the other hand for larger droplets, as $A$ approaches $N$, the relative defect imbalance decreases, and the number of defects approaches $fA$.
Then $\mu\approx 0$ and $\nu(f)\approx f$ when $A\lesssim N$, as in eq.~\eqref{Eq:nuN_t}.
We will return to this consideration in \S \ref{sec:ML} when we analyse the growth of a domain beyond its critical nucleating size.
	
	By including the theoretical estimates above in the free-energy calculation eq.~\eqref{eq:enbar}, we can derive the critical domain size, identified by the value of the area corresponding to the maximum of $\Delta F$. This yields
	\begin{equation}
		A_\text{cr} = \frac{\pi\sigma^2}{4\lambda \big(m_\text{s}(T)\,h(1-\nu(f))+2\sigma\nu(f)\big)^2 },
		\label{Eq:CriticalRadius}
	\end{equation}
	with related critical free energy barrier
	\begin{equation}
		\Delta F_\text{barr} = \frac{\pi \sigma^2}{2\lambda\big(m_\text{s}(T)\,h(1-\nu(f))+2\sigma\nu(f)\big)}.
		\label{Eq:Fc}
	\end{equation}
	
These predictions, which rely only on a single fitting parameter (the geometric factor $\lambda$ defined in \S\ref{subsec:DFEB}), will be tested against the simulation results below. It is worth mantioning that the critical domain size certainly depends on the external field $h$ triggering the reversal process. However, $A_\text{cr}$ does not diverge as $h\to 0$, as it happens in absence of defects. Any finite defect fraction $f$ induces indeed a critical area size such that the system ordering will be destroyed sooner or later, once the thermal fluctuations will generate a domain of the critical size. This is in agreement with what we noted in \S\ref{subsec:DRF} \cite{70lark,75imry}: in the presence of any finite defect fraction the system long-range ordering is eventually unstable.
	
	Once the thermal fluctuations allow the system to overcome the free energy barrier in \eqref{Eq:Fc}, the droplet invades the full domain, possibly merging with other droplets (either sub or supra critical) that might be present in the system. In the defect-free condition the velocity $v_\perp$ of expansion of a spherical droplet can be estimated by considering the Allen-Cahn approximation \cite{Allen1979}
	\begin{equation}
		v_\perp(R) = \Lambda_R\,(R_\text{cr}^{-1}-R^{-1}),
		\label{Eq:velocity}
	\end{equation}
	where $\Lambda_R$ is a temperature dependent coefficient. Since we aim also at understanding whether the presence of defects breaks the spherical symmetry of the droplets we slightly modify eq.~\eqref{Eq:velocity} to write it in terms of the domain area
	\begin{equation}
		v_\perp(A) = \Lambda_A\,\Big(A_\text{cr}^{-\frac12}-A^{-\frac12}\Big).
		\label{Eq:velarea}
	\end{equation}
	
	Expression \eqref{Eq:velarea} is expected to hold for defect-free systems. We are now interested in verifying its possible applicability in the presence of defects.
	
	\subsubsection*{Simulation results}
	
	In order to test the theoretical estimations, and to extract information on the remaining relevant parameter, we extracted the growth velocity $v_\perp$ from our numerical simulations, as the algorithm described in \S\ref{subsec:DMCA} implements the Glauber dynamics, and allows to trace how the physical time evolves across the Monte-Carlo single steps. In order to derive the growth velocity, we perform a set of simulations, all starting from the configuration in which a negative external field $h_\text{ext}=-h$ is applied on a positively-magnetized system. We fix a step $\Delta t$ for the physical time interval, and record the system configuration at each $\Delta t$. To average out rapid fluctuations, each snapshot comes in fact from the average of 10 configurations, each taken at an interval $\Delta t /10$. The value $\Delta t$ was chosen so to allow (in average) that $N$ moves were performed within each recording interval $\Delta t /10$. Next, a cluster identification algorithm is used to monitor the evolution of each single domain, and correspondent clusters (in subsequent snapshots) are associated. We remark that in our simulations we considered geometrical clusters, defined as a group of connected equally oriented spins, even though other cluster definitions could be considered, such as the physical clusters \cite{Schmitz2013}. Geometric and physical clusters have been shown to provide comparable results in nucleation theory away from the critical temperature \cite{quig21}.
	
\begin{figure}
		\centering
		\includegraphics[width=0.48\linewidth]{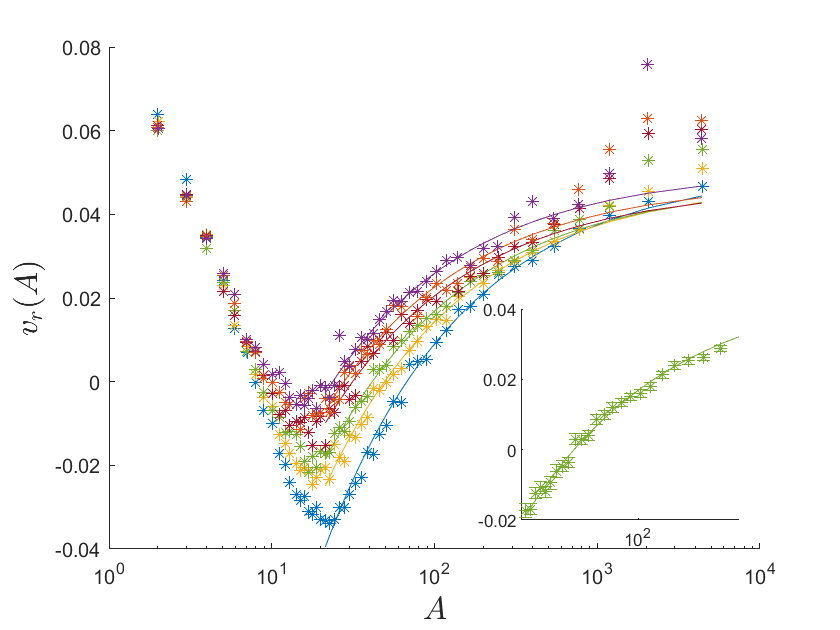} \hspace{0.02\linewidth}
		\includegraphics[width=0.48\linewidth]{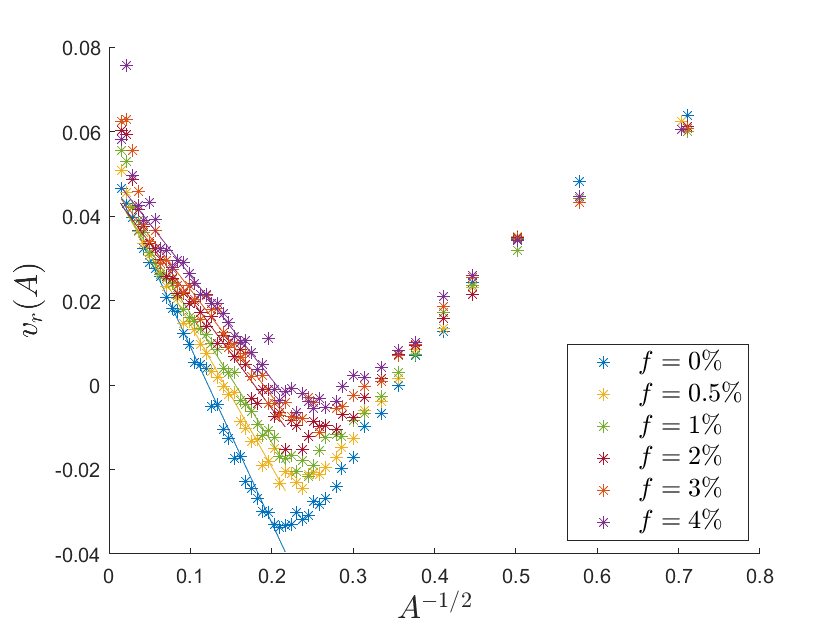}
		\caption{Domain growth velocity for a $100\times100$ system. Left: Velocity as a function of the domain area. The inset shows the typical magnitude of the error bars due to sample-to-sample fluctuations. Right: Same data, but plotted as a function of $A^{-1/2}$, as an improved test of the Allen-Cahn prediction eq.~\eqref{Eq:velarea}. The inset shows the color code (for both panels) for different defect fractions. Temperature and external field are respectively fixed at $T=0.8T_\text{c}$ and $h=0.1$.}
		\label{fig:RadialVelocity}
	\end{figure}

	In order to obtain a clean estimate of the growth velocity of a single cluster, we discarded from our statistical analysis all the snapshots which included cluster merging or cluster separation. These processes would in fact provide stochastic jumps in the growth velocity, with the jump intensity dependent on the size of the merging/separating domains. Figure~\ref{fig:RadialVelocity} shows the how the average velocity of all clusters of the same size depends on the droplet area. The same set of data is reported twice, either as a function of the defects area $A$ or as a function of $A^{-1/2}$, to better test the theoretical prediction \eqref{Eq:velarea}. For each value of the defect fraction $f$ we performed 100 simulations. Because of the presence of a free-energy barrier as in \eqref{Eq:Fc}, the simulations might include a possibly long initial period characterized by domain fluctuations, in which up to $10^3$ different domains could be traced, none of which succeeded in overcoming the free-energy barrier and therefore triggering the reversal process. Once the supra-critical domain was formed, the reversal process was quite rapid. It is not easy to provide a quantitative approximate estimate of the time a system typically spends in the fluctuation regime, as we will see that this time diverges exponentially as the external field and the defect fraction decrease.
	
	\begin{figure}
		\centering
		\includegraphics[width=0.75\linewidth]{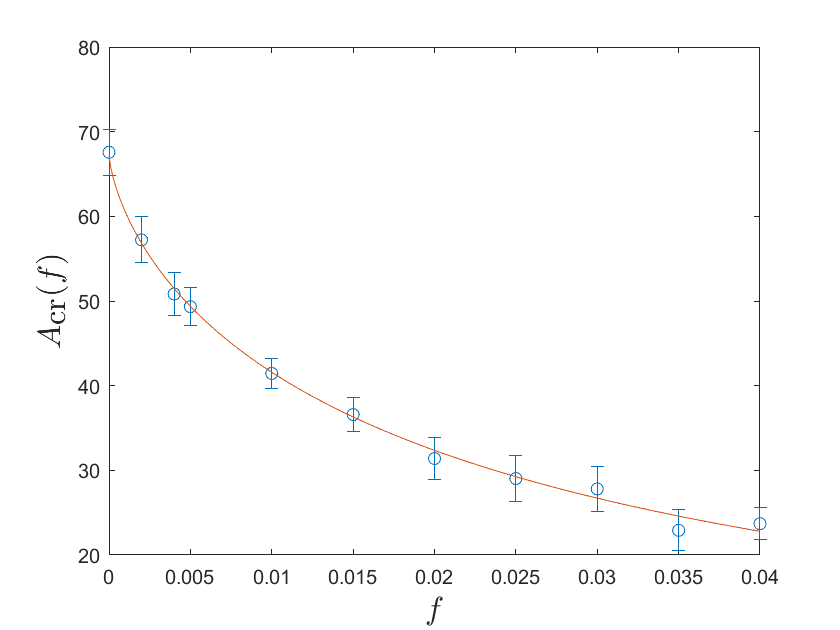}
		\caption{Critical area computed from the linear fit in the right panel of Figure \ref{fig:RadialVelocity}, as a function of the fraction of defects. The error bars reported are computed from standard error propagation theory, starting from the parameters fitted in the Allen-Cahn prediction, eq.~\eqref{Eq:velarea}. The red line shows the fit $A_\text{cr}$ vs $f$, obtained from eq.~(\ref{Eq:CriticalRadius}), with $\nu(f) = \nu_\text{opt}(f)$}
		\label{fig:CriticalArea}
	\end{figure}
	
	The structure of the velocity profile in the left panel of Fig.~\ref{fig:RadialVelocity} shows some quite different regimes. At very small domain sizes the growth velocity is positive, as this accounts for the nucleation of (eventually unstable) domains. Once the domain is generated, the average growth velocity becomes negative, because most of the domains simply revert their growth until annihilation. At a critical domain size $A_\text{cr}$ the growth velocity becomes positive and we enter the expansion regime.
	
	As our results involve averaging over large numbers of individual simulations, the question of the reliability of the averaging process is of some importance. The inset in the left panel of Fig.~\ref{fig:RadialVelocity} reports the error bars obtained for one set of data -- the velocities in the growing regime for a system with $f=1\%$ -- as a typical example. Error bars of similar magnitude are found in all other data sets. The error bars confirm that the number of replicas chosen for each value of $f$ is enough to limit the sample-to-sample variation of the growth velocity. This comes as no surprise, as the growth velocity is expected to be a quantity which depends  on the local structure around the droplet, rather than on details about the global defect distribution or concentration.
	
	In order both to check the validity of the Allen-Cahn prediction eq.~\eqref{Eq:velarea} and also to derive the most possible precise estimate of the dependence of the critical area size on the defect fraction,  in the right panel of Fig.~\ref{fig:RadialVelocity} we plot the same data, but now  as a function of $A^{-\frac12}$. This plot demonstrates  convincingly that all the curves do indeed exhibit an almost linear part, ranging from quite large domain sizes down to sizes corresponding to the minimum (negative) velocity.
	
	For each value of the defect fraction $f$, we identify the points relevant to study the domain growth (rather than the domain fluctuations) as those standing at the left of the minimum growth velocity. As discussed below, we exclude the four left-most points, corresponding to largest areas, and perform a linear regression fit in the variables $v_r$ \emph{vs.} $A^{-1/2}$. The fits performed for all values of the defect fraction provide $R^2>0.95$, with $R^2$ increasing as we lower the defect fraction. As a result we can conclude that the Allen-Cahn prediction is rather well obeyed until the droplet becomes too large. 

	In the large-droplet regime the data are biased because a different effect, related to the system topology, comes into play. Large droplets are most likely to become elongated and eventually turn into connected stripes. This changes the topology and increases the growth velocity (as demonstrated in the large-area data in the panels), because a stripe is able to increase its area without increasing its boundary, and therefore its growth is not restrained by the surface tension.
	
	Fig.~\ref{fig:CriticalArea} shows how the critical domain size $A_\text{cr}$ (as extracted from the linear regression fit just discussed) depends on the defect fraction $f$. The estimation errors are derived considering standard error propagation theory, starting from the fitted parameters from Allen-Cahn formula, eq.~\eqref{Eq:velarea}. The data points are then fitted to eq.~(\ref{Eq:CriticalRadius}), considering $\nu(f) = \nu_\text{opt}(f) = 0.5 f^{0.5}$ as discussed above. The $\chi^2$ goodness of the fit, with 9 degrees of freedom, has a $p$-value smaller than $0.01$. The theoretical prediction agrees thus remarkably well with the data extracted from our simulations. We note that the reported estimation of the critical area identified as the value at which the droplet overcomes the free energy barrier, provides a fairly stable prediction, as other possible strategies in the literature (size at which the droplet has a 50\% probability to cover the whole lattice, or at which the free energy has a maximum) provide equivalent results \cite{Ryu2010}.
	
	Qualitatively, our results demonstrate that the presence of defects enhances the growth velocity of the expanding droplets.
	This can be easily understood, as the droplet finds an easier growth direction towards positions where most negative defects are already present. The existence of this local anisotropy reduces the probability of spherical growth and as a consequence also reduces the expected value of the geometric parameter $\lambda$. Quantitatively, by using eq.~\eqref{Eq:CriticalRadius} we can extract from $A_\text{cr}$ information about the average value of the geometrical parameter $\lambda$, which results to be $\lambda=0.6$. If the droplets were rectangular, such value would correspond to a rectangle with aspect ratio of about 0.4. This indicates that the presence of defects induces anisotropic domain growth. As a consequence, stripe domains and/or jagged droplets are most likely to occur in the presence of defects.
	
	\section{Metastable Lifetime}\label{sec:ML}
	
	\subsection{General Considerations}
	\label{subsec:general}
	
	In this section we focus on the typical time taken $\bar{\tau}$ for the  magnetisation reversal transition to take place. We label this time the \textit{metastable lifetime} of the original phase.  In general the process involves  two different stages. The first stage involves the formation -- nucleation -- of a domain of critical size. The  nucleation involves a stochastic Poisson-like process.   In the second stage, droplet growth and/or coalescence complete the transition. The growth process is basically deterministic, although the coalescence involves a stochastic element, depending on the expected separation of the droplets.
	
	It is known \cite{Rikvold1994,Acharyya98,Acharyya2021} that in Ising systems there are two regimes for magnetisation reversal, which have been labelled Single-droplet (SD) and Multi-droplet (MD). This result was first derived for systems without impurities, but we will find that the basic theoretical estimates are qualitatively, although not quantitatively, robust with respect to the introduction of defects. The distinguishing feature between the two regimes is the global size of the system.  Small systems are in the SD regime, while large systems are in the MD regime.  How small  is ``small'' depends on the reversal field and the density of defects.
	
	In the SD regime, the nucleation time is considerably longer than the growth time.  The limiting factor is the nucleation of the single droplet. This then grows until the nucleating droplet  invades the whole system. The complete reversal takes place before another droplet has had time to nucleate. The growth time can be neglected.
	
	In the MD regime, by contrast, which occurs when the system is large enough, many critical droplets form before individual droplet growth can invade the whole system.   The reversal time is governed by  droplet  growth and coalescence.  The limiting factor is the time taken for two neighbouring nucleating droplets to collide.  In this circumstance it is the nucleation time which can be regarded as  negligible.
	
	It turns out that that similar but different exponential expressions govern the metastable lifetimes in the SD and MD regimes. The basic physics is as follows (see e.g.~\cite{kramers1940,Becker1935,Rikvold1994,Acharyya98,Acharyya2021}).  We first discuss the SD regime, and then the MD regime.
	
	In the SD regime, the limiting process involves the nucleation of a single droplet. This requires that  the free-energy barrier $\Delta F_{\rm barr}$ \big(see eq.~\eqref{Eq:Fc}\big)  associated with the critical droplet be overcome.  The density of droplets of excess free energy $\Delta F$ is proportional to the Boltzmann factor $\displaystyle \exp\left(-\Delta F/k_BT\right)$. Thus the density $\rho_\text{c}$ of  critical droplets, and hence the probability $p_\text{c}$ that a critical droplet is present in the system,  is proportional to $\exp \left(-\Delta F_\text{barr}/(k_\text{B}T)\right)$ \cite{Becker1935,kramers1940}. The time taken $\bar{\tau}_\text{SD}$ for such a droplet to spontaneously form will be proportional to $p_\text{c}^{-1}$. Thus in the SD regime
	\begin{equation}
		\overline{\tau}_\text{SD} (f,h)  \sim \exp \left(\frac{\Delta F_\text{barr}(f,h)}{k_\text{B} T}\right).
		\label{Eq:tauSD}
	\end{equation}
	
	In the MD regime, however, roughly speaking, the metastable lifetime is  given by  the time taken for two expanding neighbouring droplets to merge. In undefected Ising systems it has been shown that \cite{Acharyya98} that in a  $D$-dimensional system  this is given by a quantity of the order of $\rho_\text{c}^{-1/(D+1)}$. In our two-dimensional case, this leads to:
	\begin{equation}
		\bar{\tau}_\text{MD}(f,h) \sim \exp \left(\frac{\Delta F_\text{barr}(f,h)}{3 k_\text{B} T}\right).
		\label{Eq:tauMD}
	\end{equation}
	
	Strictly speaking, the expressions in eqs.~(\ref{Eq:tauSD},\ref{Eq:tauMD}) also require (different)  algebraic prefactors, dependent on $h$, $f$, and the system size $L$. However, the exponential terms in these equations are so dominant that our simulations are not able to extract precise quantitative information about these prefactors.  For this reason we only focus on the dominant terms in $\tau$. We will return to the difference between the SD and MD regimes in \S\ref{subsec:SL}.	
	
	The intuitive meaning of the metastable lifetime is clear, but a precise definition of the metastable lifetime requires a more careful examination. Indeed, when we measure the metastable lifetime and we stop the simulation, we want to be sure that the reversal has proceeded to a point that a fluctuation in the opposite direction -- involving  a reduction of the area of the droplet -- could not reverse the process. Specifically, we need a prescription of the configuration in which the system is regarded as having escaped the metastable free-energy well of a completely magnetised system in the opposite direction to the external field.
	
	In order to compare our results more easily with the previous analogous studies on defect-free systems by Rikvold and collaborators \cite{Rikvold1994}, we adopt an operational definition of the metastable lifetime $\bar{\tau}$.  This will be the time taken by the system, starting fully oriented at $m=1$,  until the magnetisation drops below a   threshold level of $m=0.7$. The criterion may seem arbitrary. An alternative similar criterion which we have used in some test simulations is that  the supracritical droplet fills  15\% of the system.   However, we find that the structural results which follow, i.e. the dependence of $\overline{\tau}$ on  defect fraction $f$ and reversal field $h$, are  robust with respect to the precise criterion for the irreversible nucleation of the spin-reversed state.  This is consistent with the results of other workers \cite{Rikvold1994}. As a result, using the operational definition is a satisfactory strategy.
	
	\subsection{Magnetic field dependence}\label{subsec:MFD}
	
\begin{figure}
		\centering
		\includegraphics[width=\linewidth]{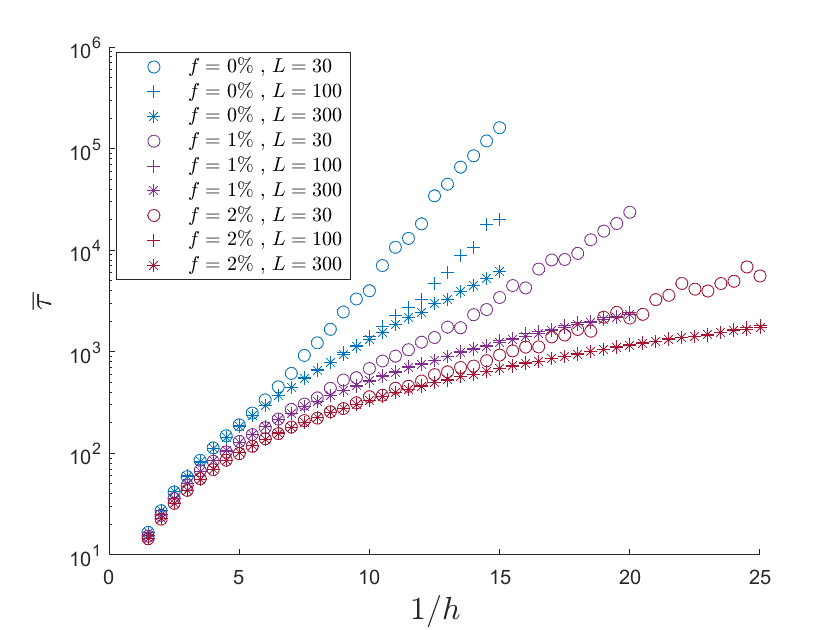}
		\caption{Metastable lifetime for a defect-free system (blue, upper points) and for systems with $f=1\%$ (yellow, middle points) and $f=2\%$ (purple, lower points), as a function of the inverse of the magnetic field. For every choice of $f$ and $h$ results are reported for different system sizes. Error bars (not shown) are on the same scale as the markers. \hfill\break}
		\label{fig:Metastable Lifetime}
	\end{figure}

	We first present an overview of simulation results of  the dependence of the average metastable lifetime $\overline{\tau}$	on the external magnetic field $h$, as defect fraction $f$ and system dimension $L$ are varied.  Some illustrative results are shown in Fig.~\ref{fig:Metastable Lifetime}. We then note some important features of these results, whose implications  we examine further in subsequent subsections.
	
	In Fig.~\ref{fig:Metastable Lifetime}, the blue points refer to defect-free simulations, and are in agreement with \cite{Rikvold1994}. In particular, by noting that the $\bar\tau$-scale is logarithmic, the observed linear dependence of $\overline{\tau}$ on $h^{-1}$  is consistent with the exponential dependence of $\overline{\tau}(\Delta F)$ given in eqs.~(\ref{Eq:tauSD}), (\ref{Eq:tauMD}), with $\Delta F_\text{barr}$ given by eq.~\eqref{Eq:Fc} with $f=0$.
	
	The purple and red points report the effect of defects. The presence of defects clearly reduces the metastable lifetime. The system is more reactive and escapes more easily from the metastable free-energy well. On this logarithmic scale the dependence $\overline{\tau} (h^{-1})$ is sublinear. The $h^{-1} \to \infty$ limit (corresponding to $h \to 0$) is unclear, especially for larger ($L=300$) systems, as the metastable lifetime data are compatible with either a slow divergence and an asymptote. But in fact, although it is not obvious here, a different plot will next suggest that for these values of $f > 0$, the metastable lifetime does tend to a finite value when the external field vanishes. We discuss this in greater detail in the next subsection.
	
	We note also that the metastable lifetime $\overline{\tau}$ exhibits some system size dependence. In all cases $\overline{\tau}$ is larger for  small systems than it is for large systems. This effect increases  as $f$  is reduced.  In Fig.~\ref{fig:Metastable Lifetime}, the $f=0$ plots show significant $L$ dependence, whereas the $f=0.02$ case exhibits a much weaker $L$ dependence.

	To sum up, Fig.~\ref{fig:Metastable Lifetime} demonstrates that different regimes exist, depending on the external field, the defect fraction and the system size. These are:
	\begin{itemize}
	\item[(a)] In the strong-field regime (left portion of the plot) $\overline{\tau}$ becomes independent of the fraction of defects and the system size and all curves collapse onto a universal $\overline{\tau}\sim 1/h$ law.
	\item[(b)] At intermediate values of the external field, the $\overline{\tau}$ \emph{vs.} $1/h$ curves change their slope, but the $\overline{\tau}$ remains basically size-independent.
	\item[(c)] If we lower the field further (right portion of the plot), data corresponding to different system sizes separate, informing us that a nucleation regime appears.
	\end{itemize}
	In Sect~\ref{subsec:SL} below we will further investigate these points. We shall identify the regimes (b),(c) as the multi-droplet (MD) and single-droplet (SD) nucleation regimes, separated by a spinodal line,  with larger $h$ and $f$ favouring the MD regime.
	
	\subsection{Zero-field Case}\label{subsec:ZFC}
	
	In the previous subsection we have speculated about the behaviour of the metastable lifetime  in the zero field limit. We first note that the zero-field case $h=0$  is an exceptional limit for all $f \ge 0$.  The metastable lifetime at finite field $h$  refers to a \emph{reversal} transition. At $f=0$, in equilibrium, the system is ferromagnetic. In the absence of a reversing magnetic field, the spins will not reverse, which corresponds to the simulation result that $\overline{\tau} \to \infty$ in the limit that the reversing field disappears. In the $ f >0$ case, as we have seen in \S \ref{subsec:DRF}, for $h=0$ the ferromagnetic state is no longer stable, but rather the equilibrium state is disordered \cite{70lark,75imry}.
	
	For $f>0$  an initially ordered state at zero field will no longer remain in this state, but rather relax to the disordered equilibrium. Thus in the $h \to 0$ limit,  the reversal transition becomes a \emph{disordering} transition.  We expect nevertheless that  the metastable lifetime associated with the limit $h \to 0$  of the \emph{reversal} transition,  and that associated with the $h=0$ \emph{disordering} transition,  will coincide.  This is the case we focus on in this section. When defects are introduced, the system is expected to escape from the ordered state in finite time even without an external field. Expression \eqref{Eq:Fc} for the free-energy barrier confirms the above expectations. Indeed, $\Delta F_\text{barr}$ increases when external field $h$ is lowered. Nevertheless, in the presence of defects it does not diverge as $h\to 0^+$. Quantitatively, from eqs.~(\ref{Eq:tauSD}) and (\ref{Eq:tauMD}), the metastable lifetime depends exponentially on the free energy barrier, while eq.~\eqref{Eq:Fc} gives an expression for the free energy barrier. It follows that $\Delta F_\text{barr}$ increases when external field $h$ is lowered, but for $f >0$, it does not diverge as $h\to 0$.
	
	For large enough systems (i.e., systems in the MD regime) we can obtain a theoretical prediction for the metastable lifetime of the disordering transition by substituting eq.~\eqref{Eq:Fc} into \eqref{Eq:tauMD} and setting $h=0$, to obtain:
	\begin{subequations}
	\begin{equation}
		\overline{\tau}(f,0) \sim \exp \left(\frac{\pi \sigma}{12k_\text{B} T\mu\lambda\nu(f)}\right).
		\label{Eq:metaA}
	\end{equation}
	
	\begin{figure}
		\centering
		\includegraphics[width=0.8\linewidth]{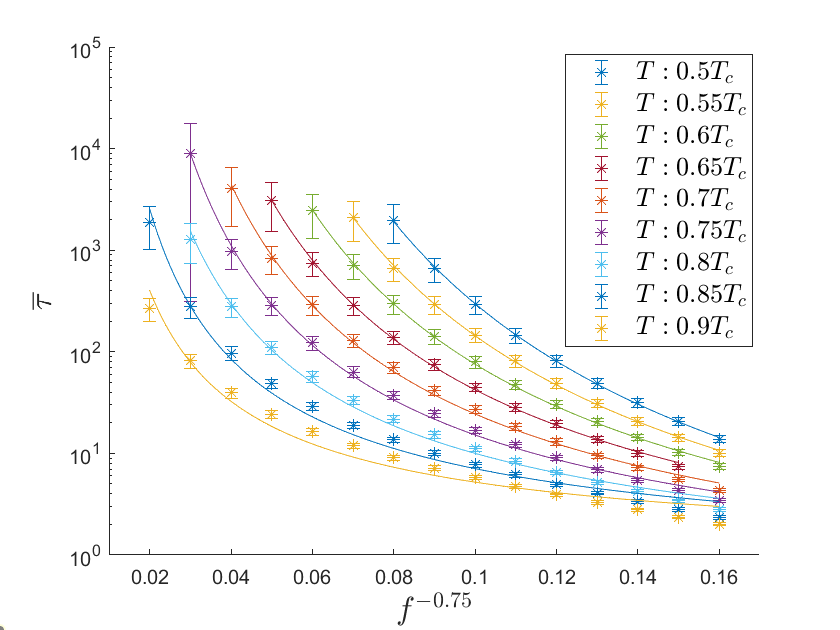}
		\ \\
		\includegraphics[width=0.8\linewidth]{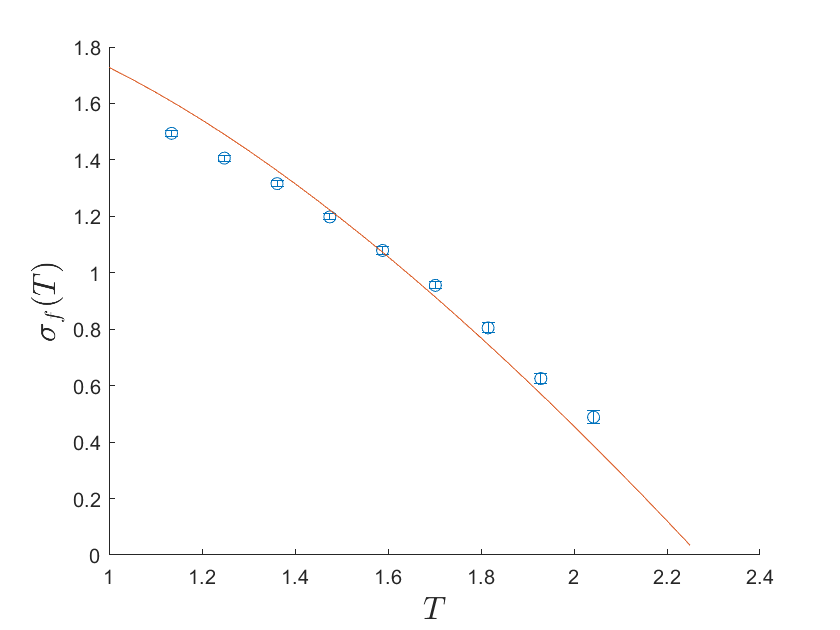}
		\caption{\textbf{Upper Panel:} Metastable lifetime as a function of the defect fraction for the displayed values of the temperature, at zero external field. For each point, we considered 400 different defect realizations for a $100\times100$ square system. \hfill\break
		\textbf{Lower Panel:} Temperature dependence of the estimated surface tension, as compared with the theoretical prediction \cite{Shneidman1999}. \hfill\break}
		\label{fig:Metastable Lifetime H0}
	\end{figure}
	
	For each value of the defect fraction $f$, we carried out simulations of the $h=0$ disordering process in 400 different systems, each containing $100\times 100$ spins. In each case the system is prepared in a perfectly ordered state. Results for the metastable lifetime $\overline{\tau}$, as a function of $f$ and $T$, including error bars, are presented in the top  panel of Fig.~\ref{fig:Metastable Lifetime H0}.	

	We also make a comparison between the simulation results and  the prediction of eq.~\eqref{Eq:metaA}. This can be written in separable form as:
	\begin{equation}
		\ln\overline{\tau}(f,0) \sim \left(\frac{\pi \sigma}{12k_\text{B} T\mu\lambda\nu(f)}\right) \approx  A(T)B(f).
		\label{Eq:metaB}
	\end{equation}
	\end{subequations}
	Equivalently, the functional form means that in principle, the curves, both as a function of  $\mu$ and as a function of $T$, should collapse onto a universal curves. The quantity $A(T)B(f)$  follows approximately from the explicit formula as follows.

	We first discuss $A(T)$. Although in principle, the surface tension $\sigma$  should include both $f$ and $T$ dependence, but we shall suppose that the principal dependence is on temperature, and the zero $f$ value can be taken for small $f$. Our simulations suggest that the other primarily  $T$-dependent quantity is the geometric factor $\lambda$.

	We now turn to $B(f)$. The droplets to be considered here are initially critical,  but have grown by a significant factor, so that they are considered when $m=0.7$, which is our criterion for an `established' droplet.  Within these droplets, we find computationally that the imbalance parameter $\mu \approx 0.5$ and is insensitive to either $T$ or $f$.   The droplets are much larger than critical, for which eq.~\eqref{Eq:CriticalNu} gives $\nu(f) \sim f^{0.5}$. But they are much much smaller than macroscopic droplets, for which eq.~\eqref{Eq:nuN_t} dictates that $\nu(f) \sim f$.  Thus we might expect that in our results, $\nu(f) \sim f^\alpha$  with  $\alpha$ somewhere between 0.5 and 1.

 	By evaluating the factors required to approximately collapse the $f$ dependent curves onto each other, the temperature dependence of the effective $\sigma(T)$ can be obtained by  combining the simulations and eq.~\eqref{Eq:metaB}. This is shown in the lower panel of Fig.~\ref{fig:Metastable Lifetime H0}. The same plot also compares the effective $\sigma$ with the defect-free computations of $\sigma(T)$ by Shneidman and coworkers \cite{Shneidman1999}, displaying impressive agreement, and justifying \textit{ex post facto} our theoretical treatment.     A best fit of the low $f$ form of $\nu(f)$ using the same set of curves yields $\alpha \approx 0.75$, i.e $\nu(f) \sim f^{0.75}$. This is of the expected form and in the expected range.
 	
 	We note also that the surface tension vanishes as the critical temperature is approached. This property can be used in complex lattice models to derive the critical temperature in absence of a theoretical exact value \cite{Muller1977}. Our estimated data for the surface tension -- albeit estimated from results from a defected model -- are qualitatively consistent with an intersection with the temperature axis close to the value predicted by Onsager's exact solution for the pure Ising model. 	 	
	
	\subsection{Spinodal Line}\label{subsec:SL}
	
	We now study the crossover between the SD and MD regimes  for magnetisation reversal \cite{Rikvold1994,Acharyya98,Acharyya2021},  extending our discussions in \S \ref{sec:I} and \ref{subsec:general}.  In pure systems (i.e., $f=0$)  this phenomenon has been studied by Rikvold et al. \cite{Rikvold1994}, who label the crossover as the ``spinodal line''.  Here we extend this study to include  defected systems. A key finding in \cite{Rikvold1994} was that there is both a microscopic and a macroscopic distinction between the SD and the MD regimes. We shall find that this classification is qualitatively, although not quantitatively, robust with respect to the introduction of defects.
	
	The microscopic distinction has been addressed in \S\ref{subsec:general} and involves the mechanism whereby the magnetic reversal takes place. The SD regime obtains when a single nucleated droplet invades the whole system. In pure systems, this is the low $L$, low $h$ regime. The opposite MD regime is dominated by droplet coalescence, and applies for higher $L$ and/or $h$. Keeping account of which is which in a simulation involves very detailed observations. Luckily there is another way.
	
	The macroscopic distinction involves the statistical distribution of reversal times over repeated simulations of systems with the same control parameters. In the SD regime, the limiting time is that required for the creation of the nucleating droplet, given by the  JMAK theory \cite{fanfoni,kolmo,avra,john}. This is a Poisson process, giving rise to an Exponential distribution with a standard deviation equal to the mean time, given by eq.~\eqref{Eq:tauSD}. In the MD  regime, by contrast, the limiting time is the time for coalescence of neighbouring droplets, given by eq.~\eqref{Eq:tauMD}. This is a Gaussian process, with a standard deviation considerably smaller than the mean reversal time. In the limit of a very large system it can be regarded as essentially deterministic.
	
	The ratio  $\rho$  of the standard deviation to the mean metastable lifetime  enables  one to distinguish between the Exponential and Gaussian distributions. In fact, the distinction is not completely sharp. Rather than a crossover line, there is in fact a fuzzy crossover region as one moves through $(L,\,H,\,f)$ space, over which $\rho$  reduces from unity (in the SD regime) to values much smaller than unity (in the MD regime). In their studies of analogous undefected systems, Rikvold et al.\ in \cite{Rikvold1994} adopted as an operational criterion for the spinodal line that $\rho_\text{c}=0.5$. Then for $\rho <0.5$, they define the system to be in the MD regime, while for $\rho>0.5$, it is in the SD regime. In our studies, we adopt the same criterion.
	
\begin{figure}
		\centering
		\includegraphics[width=\linewidth]{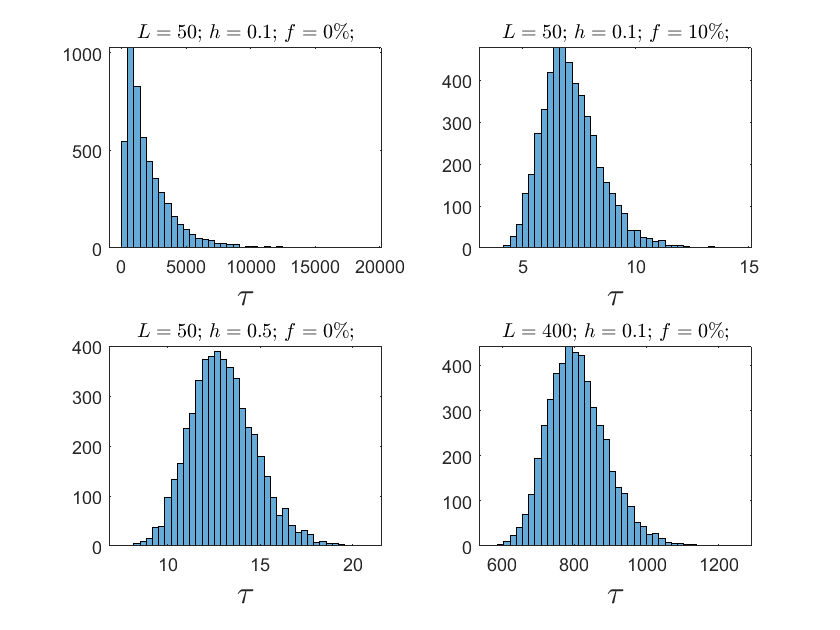}
		\caption{Distribution of metastable lifetimes $\tau$  for some example systems.  System size $L$,  defect fraction $f$ and external field $h$ shown above each subfigure.  Simulation temperature: $T=0.8~T_\text{c}$ with  $T_\text{c}$  the  Curie temperature for the defect-free system.}
		\label{fig:MeanDistribution}
	\end{figure}

	Illustrative examples of how this works in practice are shown in Fig.~\ref{fig:MeanDistribution}. In this figure, three of the four panels exhibit a Gaussian-like structure and correspond to MD reversal. By contrast the top left hand panel clearly possesses a Poissonian-like structure, and  corresponds to SD reversal.

	It is useful to carry out  a pairwise comparison between the top left and each other panel in Fig.~\ref{fig:MeanDistribution}.   Comparing the top left and bottom left panels, $h$ increases. Comparing   the top left  and top right panels,  $f$  increases. Comparing the top left and bottom right panels,  $L$ increases.  In each case,  increasing  a single  control parameter,  while keeping other parameters constant,   shifts the system from the SD to the MD regime, and significantly accelerates the reversal process.
	
	We draw the reader's attention to two further features of the distributions shown in Fig.~\ref{fig:MeanDistribution}. Firstly, the kurtosis in the shape of the distribution in the top right panel illustrates the actual fuzziness of the transition.  Here the transition from Poissonian to Gaussian is almost but not quite complete.  By  the $\rho_\text{c}=0.5$ criterion, however, this case falls clearly into the MD regime.
	
	Secondly,  we look now not at the distribution  shapes, but merely at the magnitudes involved.  Comparing the top left panel with either the top right panel (involving an increase in $f$) or the bottom left panel (involving an increase in $h$) shows a dramatic reduction in the metastable lifetime. By contrast, comparing the top left and the bottom right panel (involving only an increase in system dimension  $L$), the mean metastable lifetime is left largely unchanged. Although there is insufficient evidence from these figures alone, it looks as though the behaviour of the system is in some sense saturating as the system size is increased. This remark is confirmed also by the data reported in Fig.~\ref{fig:Metastable Lifetime}, which show a remarkable reduction in the metastable lifetime when we move from small (SD) to medium (MD) systems, but a much minor effect when the system size is further increased (compare the data for systems with $L=100$ and $L=300$).
	
\begin{figure}
		\centering
		\includegraphics[width=0.8\linewidth]{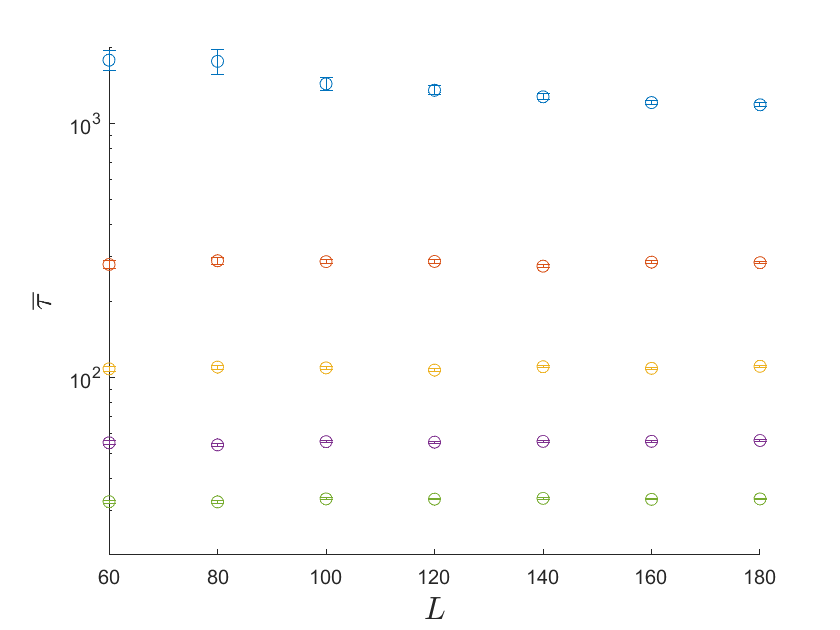}
		\ \\
		\includegraphics[width=0.8\linewidth]{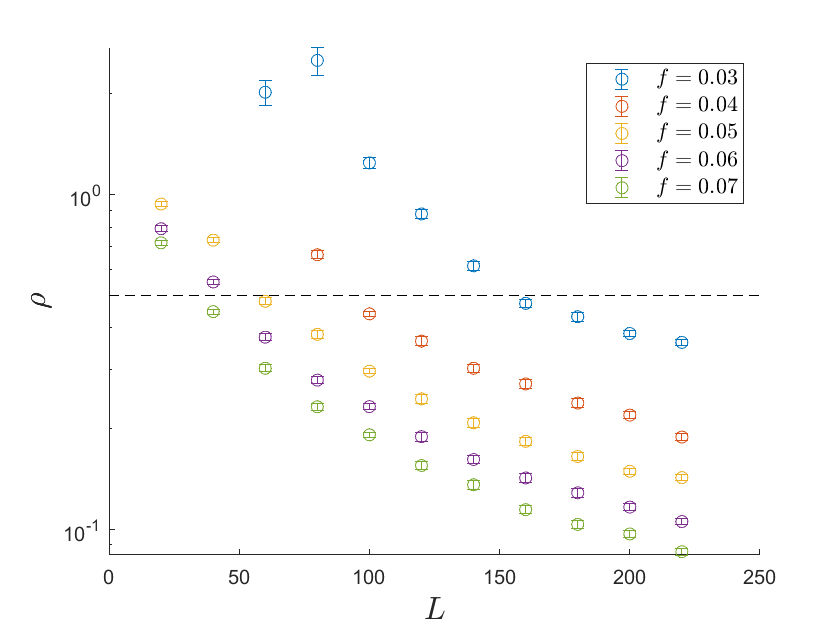}
		\caption{ Metastable lifetime $\overline{\tau}$ (upper panel)  and relative standard deviation $\rho$ (lower panel) as a function of  system size for several values of defect fraction $f$.  $h=0$, $T=0.8 T_\text{c}$. Legend (both plots): top to bottom   $f=0.03$ (blue), $f=0.04$ (red) $f=0.05$ (orange), $f=0.06$ (purple) $f=0.07$ (green). \hfill\break}
		\label{fig:RSDH0a}
	\end{figure}

	We now turn our attention to the SD-MD transition in the $h=0$  case for finite $f$.  Some features of this have already been discussed in \S\ref{subsec:ZFC}, in which we varied the temperature $T$ and defect fraction $f$, at constant system size $L=100$.  In this next set of simulations, we keep $T=0.8T_\text{c}$  constant and allow $f$ and $L$ to vary. The quantities of interest are the mean metastable lifetime $\overline{\tau}$, and the ratio of the standard deviation to the mean of the metastable lifetime distribution $\rho$. Key  elements of these results are shown in Fig.~\ref{fig:RSDH0a}.
	
	The upper panel of Fig.~\ref{fig:RSDH0a}  presents the dependence of the mean metastable lifetime $\overline{\tau}$ as a function of  $L$ for different values of $f$. At constant  $f$, $\overline{\tau}$ is rather insensitive to changes in $L$.  In the lower panel, we plot the dependence of  $\rho(L)$. Here, by contrast, the system size effects are strong.  The dotted line indicates $\rho_\text{c}=0.5$, the SD-MD transition. The critical size $L_\text{c}(f)$ is a decreasing function of $f$; for $f=0.03$, this transition occurs at $L_\text{c} \approx 160$, while by $f=0.07$, the critical size has reduced to $L_\text{c}\approx 40$.   Thus  as the defect fraction $f$ decreases, larger systems are required for an essentially deterministic transition to occur.
	
	The JMAK theory allows us again to draw a theoretical prediction about the system sizes at which the nucleation process is expected to enter the MD regime. The nucleation rate $I$ (that is, the number of droplets of the critical size nucleated per unit time in the system) is inversely proportional to the exponential factor in Eq.~\eqref{Eq:tauSD}. The average distance $R_0$ between such critical droplets is inversely proportional to $I$, and thus in turn proportional to the exponential factor in Eq.~\eqref{Eq:tauSD}. The phase transition can be expected to be a SD or a MD process when respectively $R_0\ll L$, or $R_0\gg L$. It is therefore natural to expect that the location of the spinodal line defined above qualitatively coincides with the condition $R_0\approx L$, so that in view of Eq.~\eqref{Eq:tauSD} we put forward the theoretical prediction
	\begin{equation}
		L_\text{SL}\sim \exp \left(\frac{\Delta F_\text{barr}(f,h)}{3k_\text{B} T}\right),
		\label{eq:LSL}
	\end{equation}
	where again we are dropping possible algebraic pre-factors to capture the dominant free-energy and temperature contribution.
	
\begin{figure}
		\centering
		\includegraphics[width=\linewidth]{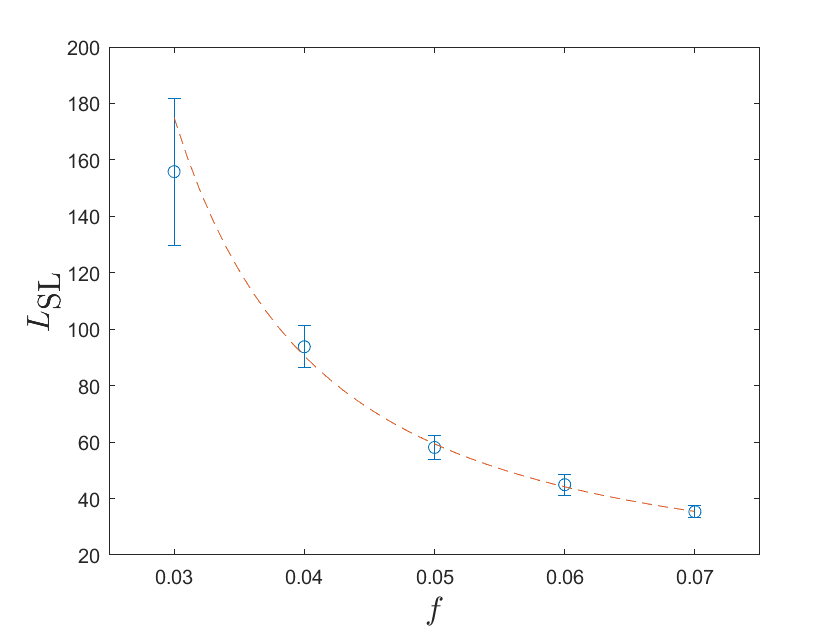}
		\caption{Critical system size $L_\text{SL}$ at the spinodal line as a function of the defect fraction in the absence of external field. Error bars are obtained from standard error propagation theory as explained in the text. The fit is obtained through Eq.~\eqref{eq:LSL} with $\nu(f)\sim f^{0.75}$ as discussed in \S\ref{subsec:ZFC}.}
		\label{fig:RSDH0}
	\end{figure}

	The height of the free-energy barrier in Eq.~\eqref{eq:LSL} decreases when either the external field or the defect fraction increases. For this motivation, we again focus first on the $h=0$ case, which provides a limiting value for $L_\text{SL}$, that is a critical system size below which the disordering  transition will certainly occur in the SD regime, at any finite value of the external field.	Figure~\ref{fig:RSDH0} shows how $L_{SL}\big|_{h=0}$ depends on $f$. The dotted line shows the fit to eq.~\eqref{eq:LSL}, which provides a $p$ of about $0.15$ for the $\chi^2$ goodness of the fit test with 3 degrees of freedom. The errors in the measured values are obtained from standard error propagation theory, starting from the estimation errors of the precise value at which the curve $\rho(L)$ crosses the threshold value 0.5. The simulation points are displayed only for $f\geq 3\%$, as lower values require extremely large systems to find the transition to the MD regime in the absence of magnetic field. To make a specific example, for $f=1\%$, the expected value is $L_\text{SL}\big|_{h=0}\simeq 10^6$.
	
	\begin{figure}
		\centering
		\includegraphics[width=0.8\linewidth]{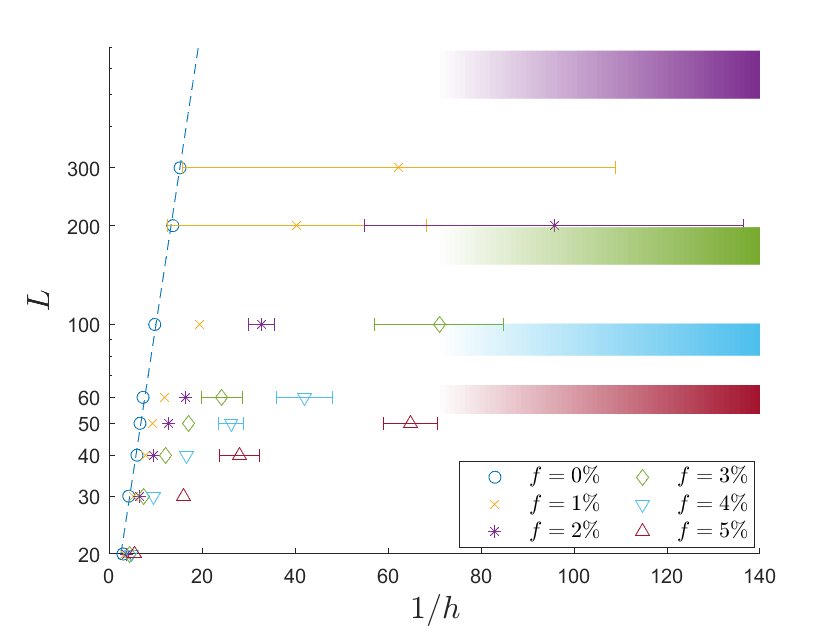}
		\caption{Phase diagram summarizing the location of the spinodal line, as a function of the system size $L$ and the magnetic field $h$, for different values of the defect fraction $f$. Each color represents a value of the defect fraction $f$, as specified in the inset. For each defect fraction and each value of the system size $L$ we have located the value of the critical $h$ at which the spinodal parameter $\rho$ crosses the threshold value $\rho_\text{c}=0.5$. The critical values of $h$ are reported with horizontal error bars, as different realizations of the same defect fraction undergo the transition at somewhat different values of the external field. For each value of $f$ we also report the limiting asymptote, which represents the system size at which the transition occurs for $h=0$. These asymptotic values and error intervals coincide with those reported in Figure~\ref{fig:RSDH0}. The left-most series of data corresponds to a defect-free simulation, and agrees with \cite{Rikvold1994}.\hfill\break}
		\label{fig:SL}
	\end{figure}
	
	Figure~\ref{fig:SL} summarizes the results we obtained for the spinodal line in the presence of both an external field, and quenched defects. For the reported values of the system size $L$ and defect fraction $f$, we varied the external field $h$ while monitoring the ratio $\rho$ between the standard deviation and the mean value of the metastable lifetime $\bar\tau(L,f,h)$. The critical (spinodal) value of the external field was then identified as the value such that the ratio $\rho$ crossed the critical value $\rho_\text{c}=0.5$. To derive a reliable estimate of the critical magnetic field, several replicas of the system were built up (with defects initialized in different positions), and the error bars in Fig.~\ref{fig:SL} report the standard deviation of the critical value of $h$ obtained for different replicas. Due to sample to sample fluctuations, systems with more quenched defects require averaging over more replicas to obtain reliable estimates. In order to fix the number of replicas required for a stable average, we set a threshold for the error bars in Fig.~\ref{fig:SL}, and in particular we request that errors must not exceed the $20\%$ of the estimated value for the reported spinodal point. As an example, for the larger defect fraction reported ($f = 5\%$) up to $16.000$ different defect realizations were to be analysed. Only exception to this rule regards the systems with $L \geq 200$ for which, due to a rather long computer simulation time, a lower amount of replicas were considered: 200 for $f=1\%$, and 1000 for $f=2\%$. The bars displayed in the right part of Fig.~\ref{fig:SL} report the knowledge (derived from Fig.~\ref{fig:RSDH0}) of the low-$h$ behavior of $L_\text{SL}$, with the width of the bands reporting the error bars in Fig.~\ref{fig:RSDH0}.

	The information drawn from Figs.~\ref{fig:RSDH0} and \ref{fig:SL} confirm that the location of the spinodal line (that is, the size at which the system is sufficiently \emph{large} to exhibit deterministic response) depends strongly on the defect density. All data evidence a linear dependence of $L_\text{SL}$ on $1/h$ for large values of the external field (left part of the plot). In the absence of defects (light blue circles, coherent with data reported in \cite{Rikvold1994}) the linear regime extends also to intermediate and low values of $h$. On the contrary, the presence of defects gives rise to a saturation $L_\text{SL}(f)$ such that the system behaves deterministically as long as $L>L_\text{SL}(f)$, whatever the external field.
	
	\section{Discussion and Conclusions}
	
	We have analyzed the time dependence of magnetic relaxation in a two-dimensional Ising system. The extra feature of work as compared to previous work  is the variable fraction of defects  quenched into the system. The specific focus of interest is the manner in which an initially aligned system  escapes from a metastable aligned state either when it is subject to an opposite external field, or alternatively in zero field when the defects cause the equilibrium state to be disordered.
	
	Previous studies have determined that for undefected Ising systems, the classical JMAK theory is an excellent starting point to derive theoretical predictions for the free energy barriers and the related metastable lifetime of decaying phases. Our work shows that this remains true, even when defects are introduced into the system. In Section~\ref{subsec:DRF} we make theoretical predictions for the dependence of the critical droplet area and the related free energy barrier on the quenched disorder properties (see eqs.~\eqref{Eq:CriticalRadius}, \eqref{Eq:Fc}). These  predictions depend on a number of factors, including the defect imbalance,  the total number of defects enclosed within the droplet, and its geometrical shape.

	Then, in order to reduce the number of fitting parameters before comparing the theoretical predictions with the simulation outcomes, we use a probabilistic argument to identify the defect structure within the location where the nucleating droplet is most probably located,  as shown in Fig.~\ref{fig:estimate}. This leaves the aspect ratio of the nucleating droplet as a unique fitting parameter. Finally, to complete the theoretical derivation, we adapt the Allen-Cahn approximation to predict the growth velocity of each individual droplet, either above or below the critical size, noting that below-critical droplets have a negative growth velocity.
	
	The results of a detailed set of simulations, shown in the lower plot in Fig.~\ref{fig:RadialVelocity}, confirm that the critical nucleating area depends on the defect fraction in the manner predicted by our theory.

	The single fitting parameter for our theory is the geometrical droplet factor $\lambda$ of the nucleating droplet, corresponding to its aspect  ratio. For the range of $f$ for which  we have performed simulations,  we find that $\lambda$ takes a value of about 0.6.  Some caution is associated with this value, however. In a pure system, $f=0$,  subject to thermal fluctuations, it might be expected that a nucleating droplet would be   circular, i.e.  $\lambda = 1$.  The  square lattice anisotropy replaces the circle by a  square, with $\displaystyle \lambda ={\pi}/{4}\approx 0.79$. The presence of defects thus reduces $\lambda$  further. Intuitively,  some reduction is not unexpected, and follows from  the randomness in the defect locations. We postpone further investigation of this phenomenon to a later study.
	
	The JMAK theory also provides a framework within which we are able to make theoretical predictions concerning the average lifetime $\overline{\tau}$ of the metastable phases. In Section~\ref{sec:ML} we test those predictions against the simulation results. We confirm the theoretical expectation that the presence of quenched randomness strongly reduces the so-called metastable lifetime, as shown in Figure~\ref{fig:Metastable Lifetime H0}. An important result, which is derived theoretically and confirmed computationally, is that in the presence of defects, by contrast with the zero-defect case, the metastable lifetime in the absence of a field remains finite. These findings could potentially be of relevance to experimental research in fast switching devices, as noted in recent articles \cite{Mustafa2021,Heliyon2018}. These devices are of major importance in magnetic recording technologies, where magnetic systems with high sensitivity to external fields, and able to change  their magnetized state rapidly, are a necessity for fast recording devices.
	
	A detailed computational study in \S\ref{subsec:ZFC} of the dependence of the metastable lifetime associated with the disordering transition (i.e.~$h=0$), varying both temperature $T$ and defect fraction $f$ has been performed. The dependence on both $f$ and $T$ fall on universal curves with suitable scaling. The theory also predicts this universality. The temperature dependence given by theory and simulations is in good agreement, and is related to the temperature dependence of  the surface tension associated with boundaries between regions of opposite orientation.  However,  the defect fraction dependence  presents a puzzle which we have not so far been able to resolve.
	
	In Section~\ref{subsec:SL} we focus on the character of the reversal transition, and specifically the transition between single droplet (SD) and multi-droplet (MD) reversal behaviour. In the former case the transition is nucleated by a single droplet which grows to invade the whole system. In the latter case, the more or less simultaneous nucleation of many droplets is followed by droplet coalescence in such a way that the system is invaded by the new phase. The separation line -- the so-called ``spinodal'' -- is system size dependent, but the critical size depends on $T,h$ and $f$.  A signature of this transition is the statistical behaviour of the distribution of metastable lifetimes when the same system is reproduced many times. The SD case presents a Poisson distribution, while the MD case is Gaussian. In this latter case, the ratio of the standard deviation to the mean tends to zero for large systems, indicating a process which for large systems is essentially deterministic. Fig.~\ref{fig:SL} summarizes the information we derive to position the spinodal line in the presence of both quenched disorder and an external magnetic field.
	
	The results presented in this work can be extended in several directions. Firstly, a study of spin reversal in similar random-bond and random-field Ising models enables the statistical properties of Barkhausen noise to be reproduced \cite{metra}. An obvious question is whether the present system too mimics the Barkhausen effect. Secondly, it might be possible to introduce Monte-Carlo moves involving spin swapping. This would enable a relaxation of the postulate of fixed  quenched defects.

	Alternatively spin swapping, rather than spin reversal, would enable the modelling of a phase separating lattice gas. This  system would exhibit coarsening, rather than  phase invasion,  and so the introduction of defects might mimic two-component phase separation in a  porous network. An extension of this modified model might also allow 0-spin defects, much in the spirit of some recent works \cite{quig21}. Finally we might reasonably ask to what extent our results are robust with respect to increasing the system dimension. For example in higher dimensions, what kind of internal structure in the defect spins is not only quantitatively important, but qualitatively changes the reversal properties?
	
	\section*{Acknowledgments}
	
	TJS is grateful to the Politecnico di Milano for hospitality during the period of this study.
	
	\appendix
	
	\renewcommand{\theequation}{A.\arabic{equation}}
	\setcounter{equation}{0}
	
	\section*{Appendix. Probability of finding defects in domains}\label{app:prob}
	
	This appendix calculates some quantities of relevance in \S\ref{subsec:CSGV}.  We consider a domain composed of $A$ spin sites.  Each site has an equal probability $f/2$ of hosting a negative or a positive defect. The probability of finding $(n^+,n^-)$ defects in the domain is given by the multinomial expression:
	\begin{equation}
		P_1(A, f, n^+,n^-)=\frac{A!}{n^+!\,n^-!\, (A-n^+-n^-)!}\left(\frac{f}{2}\right)^{n^++n^-}(1-f)^{A-n^+-n^-}.
	\end{equation}
	Then define  $P_2$  to be the probability of  finding a specific number $n_\text{T}=n^++n^-$ of defects, and $P_3$ to be the probability of finding a specific imbalance $\Delta n=n^--n^+$ in this domain.  $P_2, P_3$ are then calculated by summing over the probabilities $P_1$ subject to the relevant constraints:
	\begin{subequations}
		\begin{align}
			P_2(A, f, n_\text{T}) &= \sum_{j=-n_\text{T}}^{n_\text{T}}P_1\big(A, f,\tfrac12(n_\text{T}+j),\tfrac12(n_\text{T}-j)\big);\\
			P_3(A, f, \Delta n) &= \sum_{j=\Delta n}^{A}P_1\big(A, f,\tfrac12(j-\Delta n),\tfrac12(j+\Delta n)\big).
		\end{align}
	\end{subequations}
	
	We now consider that the domain could be placed in $K$ different independent locations. In an available total area of $L^2$ spins, we can choose $K=L^2/A$ different possible independent locations for the nucleating droplet.
	We aim at tracing the choice (among the $K$ possible ones) which has the largest value of $\Delta n$. In such optimal choice the surface cost to start a reversal process (estimated in eq.~\eqref{eq:fsurf}) is minimized. Therefore, that domain identifies the optimal location to place the droplet to originate the reversal process. We first compute the probability $P_4$ of finding am imbalance \emph{smaller than} $\Delta n$ in a domain of $A$ spins
	\begin{equation}
		P_4(A, f, \Delta n)=\sum_{j=-A}^{\Delta n -1}P_3(A, f,j).
	\end{equation}
	Now, the probability that, among $K$ independent possible droplet locations, the optimal one (in terms of $\Delta n$) has a defect imbalance precisely equal to $\Delta n$ is given by
	\begin{equation}
		P_5(A, f, K, \Delta n)=P_4(A, f, \Delta n+1)^K-P_4(A, f, \Delta n)^K.
	\end{equation}
	The above calculations allows us to compute the expected value $\Delta n_\text{exp}$ of the random variable $\Delta n_\text{opt}$, obtained as the optimal (maximum) $\Delta n$ value among $K$ realizations of a domain in which both positive and negative defects have a probability $f/2$ to occur. This is given by
	\begin{equation}\label{eq:delnopt}
		\Delta n\big|_\text{exp}=\sum_{\Delta n=-A}^A \Delta n \, P_5(A, f, K, \Delta n).
	\end{equation}
	Finally, it is also possible to compute what is the expected value of the total number of defects in the domain realization which optimizes $\Delta n$. This is obtained through the Bayesian probability
	\begin{equation}
		P_6(A, f, n_\text{T}\,|\, \Delta n)=\frac{\text{Prob}\,(\Delta n\,|\,n_\text{T})\;P_2(A, f, n_\text{T})}{P_3(A, f, \Delta n)}=
		\frac{1}{2^{n_\text{T}}}\binom{n_\text{T}}{\frac{n_\text{T}+\Delta n}{2}}\;\frac{P_2(A, f, n_\text{T})}{P_3(A, f, \Delta n)}.
	\end{equation}
	
	Figure~\ref{fig:estimate} in the main text illustrates the results of the calculations performed with the parameter values ($A$ and $K$) extracted from the simulations in the main text. In them different (yet typically small) defect fractions $f$ are considered.
	
	\bibliography{Bibliography}
	\bibliographystyle{elsarticle-num}

\end{document}